\documentclass{article}
\usepackage{spconf,amsmath,graphicx,multirow,array}
\usepackage{epstopdf}
\usepackage{amssymb}
\usepackage{caption}
\usepackage{subfigure}

\title{Speakerfilter-Pro: an improved target speaker extractor combines the time domain and frequency domain}
\name{Shulin He*, Hao Li*, Xueliang Zhang\thanks{* These authors contributed equally to this work.}}
\address{College of Computer Science, Inner Mongolia University, China\\\texttt{heshulin@mail.imu.edu.cn,lihao.0214@163.com,cszxl@imu.edu.cn}
}

\begin{document}
%
\maketitle
\begin{abstract}
This paper introduces an improved target speaker extractor, referred to as Speakerfilter-Pro, based on our previous Speakerfilter model. 
The Speakerfilter uses a bi-direction gated recurrent unit (BGRU) module to characterize the target speaker from anchor speech and use a convolutional recurrent network (CRN) module to separate the target speech from a noisy signal.
Different from the Speakerfilter, the Speakerfilter-Pro sticks a WaveUNet module in the beginning and the ending, respectively. The WaveUNet has been proven to have a better ability to perform speech separation in the time domain. In order to extract the target speaker information better, the complex spectrum instead of the magnitude spectrum is utilized as the input feature for the CRN module.
Experiments are conducted on the two-speaker dataset (WSJ0-mix2) which is widely used for speaker extraction. 
The systematic evaluation shows that the Speakerfilter-Pro outperforms the Speakerfilter and other baselines, and achieves a signal-to-distortion ratio (SDR) of 14.95 dB.

\end{abstract}
\section{Introduction}
\label{sec:intro}
The human brain can focus auditory attention on specific sounds by shielding the sound background in the presence of multiple speakers and background noise. This is the so-called cocktail party problem \cite{getzmann2017switching}.
Recently, deep learning-based speech separation approaches for solving the cocktail party problem have attracted a considerable attention. 
Some of the representative advances include deep clustering \cite{hershey2016deep}, deep attractor network \cite{chen2017deep}, and permutation invariant training \cite{yu2017permutation}. More recently, a convolutional time-domain audio separation network (Conv-TasNet) \cite{luo2019conv} has been proposed and obtained a great separation performance improvement. 

Lots of researchers separate all signals from the mixture signal to deal with the cocktail problem.
However, in some practical situations, such as smart personal devices, we may be interested in recovering a target speaker while reducing background noise and other interfering speakers, called target speaker extraction.

Speech extraction exploits an anchor clue to identify and extracts the target speaker from a mixture signal. 
Recently, research on speaker extraction has gained increasing attention, as it naturally avoids the global permutation ambiguity issue and does not require knowing the number of sources in the mixtures. 
Several different frameworks based on frequency domain have been developed in recent years, including TENet \cite{li2019tenet}, SpeakerBeam \cite{vzmolikova2019speakerbeam}, VoiceFilter \cite{wang2018voicefilter} and Speakerfilter \cite{he2020speakerfilter}. 
There are also some successful frameworks in the time domain, such as TD-speakerbeam \cite{delcroix2020improving} for multi-channel speaker extraction.

Inspired by the better performances obtained in the time domain than frequency domain in the audio separation task, we propose an improved target speaker extractor, referred to as Speakerfilter-Pro, based on our previous Speakerfilter \cite{he2020speakerfilter} model. 
Different from the Speakerfilter, the Speakerfilter-Pro adds a WaveUNet \cite{stoller2018wave} module in the beginning and the ending, respectively. The WaveUNet is based on the time domain has been proven that it has a better ability in speech separation. We still use the convolutional recurrent network (GCRN) to characterize the target speaker information. To extract the target speech better, the complex spectrum instead of the magnitude spectrum is used as the input feature for the GCRN module.

Compared with our previous work, there has two contributions. 1) The WaveUNet is used to separate the target speech better. 2) The complex spectrum feature is employed to make better use of the anchor speech. Through the experiments, we find that the proposed method outperforms Speakerfilter and other baselines, and we also find that WaveUNet and the complex spectrum feature can boost each other. 

In this paper, we first explain how this work relates to our previous work (Section \ref{sec:Review of the previous work}). Then, we describe the proposed Speakerfilter-Pro (Section \ref{sec:Speakerfilter-Pro}). Section \ref{sec:experiment} describes the experimental setup and result. We conclude this paper in Section \ref{sec:conclusions}.

\section{Previous work}
\label{sec:Review of the previous work}

\subsection{Problem formulation}
\label{ssec:Problem formulation}
The purpose of speaker extraction is to extract the target speech from an observed mixture signal. The work takes anchor speech as auxiliary information. The mixture signal can be written as:
\begin{equation}
	y=s_t+s_i,
\end{equation}
where $s_t$ and $s_i$ are the target speech signal and the interfering signal, respectively. The interfering signal may comprise the speech signal from other speakers and background noise. In this paper we only consider the interfering speech. 
The speaker extractor can be defined as:
\begin{equation}
	\hat{s}_t=f(y, s_a),
\end{equation}
where $f$ is the transformation carried out by Speakerfilter-Pro, $\hat{s}_t$ is the estimated target speech, and $s_a$ is the anchor speech.


\subsection{Speakerfilter system}
\label{ssec:Speakerfilter system}

The proposed method is based on our previous work \cite{he2020speakerfilter}, and here we give a brief introduction of the Speakerfilter.
The Speakerfilter system takes two speech signals as input: noisy speech and anchor speech. 
A bi-direction gated recurrent unit (BGRU) is used to encode the anchor speech and average the outputs of BGRU in the time dimension to produce the anchor-vector. 
This anchor-vector can be viewed as the identity vector of the target speaker. Then, it is stacked to each frame of the mixture magnitude spectrum. 
We use a gated convolutional network (GCNN) to encode the anchor vector into each layer of the speech separator built by a GCRN. The GCRN is used to extract the target speech from noisy speech, with the help of the anchor-vector.

\section{Speakerfilter-Pro}
\label{sec:Speakerfilter-Pro}

\subsection{WaveUNet}
\label{ssec:WaveUNet}

Recently, performs speech separation in the time domain is very attractive. Researchers are also proven that the speech separation performs better in the time domain than in the frequency domain \cite{delcroix2020improving}. 

\begin{figure}[h]
  \centering
  \includegraphics[width=8.5cm]{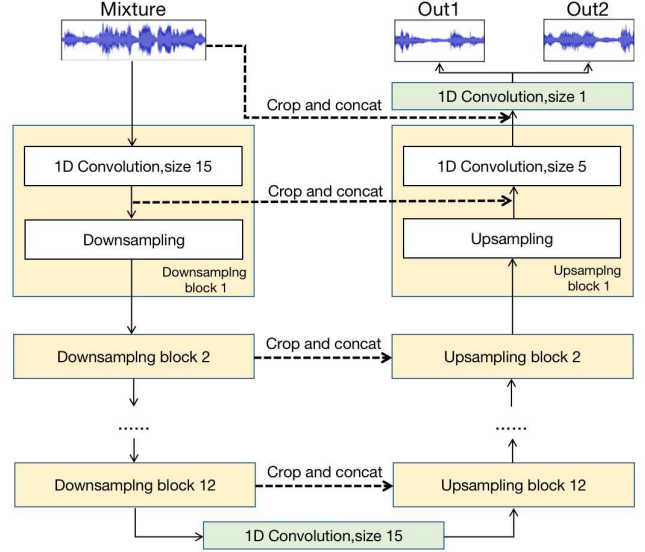}
\caption{WaveUNet structure}
\label{fig:waveunet}
\end{figure}

The WaveUNet \cite{stoller2018wave} is the most commonly used model in the time-domain. The structure is shown in Fig. \ref{fig:waveunet}. It computes an increasing number of higher-level features on coarser time scales using downsampling (DS) blocks. 
These features are combined with the earlier computed local, high-resolution features using upsampling (US) blocks, yielding multi-scale features to make predictions. 

\subsection{Model topology}
\label{Model topology}
Inspired by the WaveUNet's powerful speech separation capabilities, we propose an improved target speaker extractor, referred to as Speakerfilter-Pro, based on our previous work \cite{he2020speakerfilter}.
Different from the Speakerfilter, the Speakerfilter-Pro add a WaveUNet module in the beginning and the ending, respectively. Second, in order to encode the target speaker better, the complex spectrum instead of the magnitude spectrum is used as the input feature for the GCRN module. 

\begin{figure*}[ht]
	\centering
	\includegraphics[width=1.0\textwidth]{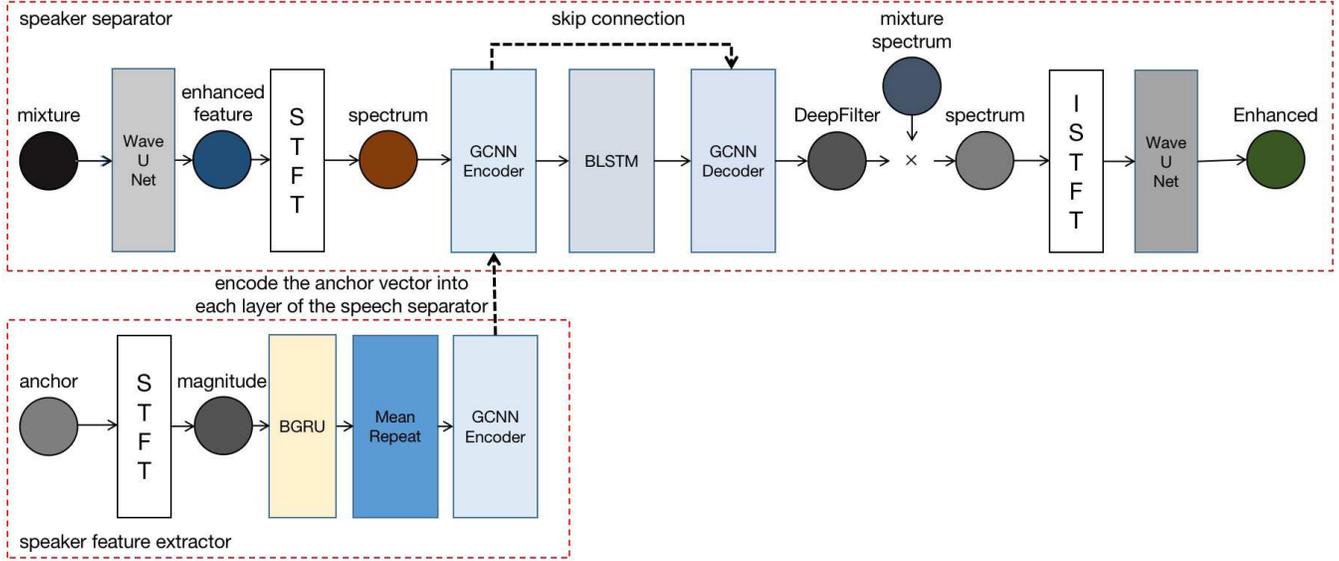}
	\caption{Overview of the Speakerfilter-Pro system.}
	\label{fig:overview}
\end{figure*}
The structure of the Speakerfilter-Pro is shown in Fig. \ref{fig:overview}.
First, the time-domain mixture signal is fed into the beginning WaveUNet.
After obtaining the output of the WaveUNet, the short-time Fourier transform (STFT) is used to transfer the time domain signal to the frequency domain.
The real and the image part of the spectrum (RI) then stack with the anchor-vector as the input to the GCRN, where the target speaker-vector is obtained by the speaker feature extractor.
With an encoder-decoder architecture, the CRN encodes the input features into a higher-dimensional latent space and then models the latent feature vectors' sequence via two bi-direction long short-term memory (Bi-LSTM) layers. 
The output sequence of the Bi-LSTM layers is subsequently converted back to the original input size by the decoder. 
To improve the flow of information and gradients throughout the network, skip connections are used to concatenate each encoder layer's output to the input of the corresponding decoder layer. 
After GCRN, the Deepfilter \cite{mack2019deep} is utilized to recover the target speech better. 
Deepfilter obtains enhanced speech $\hat{X}$ from mixture speech $X$ by applying a complex filter:
\begin{equation}
	\hat{X}(n,k) = \sum_{i=-I}^I\sum_{l=-L}^LH^*_{n,k}(l+L,i+I)\cdot X(n-l,k-i)
\end{equation}
where $2 \cdot L + 1$ is the filter dimension in time-frame direction, $2 \cdot I + 1$ in frequency direction and $H^*$ is the complex conjugated 2D filter of TF bin (n, k). In our experiments, we set $L$ = 2 and $I$ = 1 resulting in a filter dimension of (5, 3). 
Finally, the WaveUNet is used in the last part to improve the performance further.

For speaker feature extractor, the magnitude spectrum of anchor speech in frequency-domain is used as the inputs. We firstly encode the anchor speech into 129-dimensional features by using BGRU and average the outputs of BGRU in the time dimension. 
This 129-dimensional anchor vector can be viewed as the identity vector of the target speaker. 
Then, it is stacked to each frame of the mixture feature. To utilize the speaker information effectively, we propose to use CNN to encode the anchor vector into each layer of CRN, as shown in Fig. \ref{fig:overview}.

The whole modules are training jointly. The training loss is defined in the frequency domain.
The loss function is mean squared error (MSE), which is given by:
\begin{equation}
	L_{mse} = \frac{1}{N*K}\sum_{n=1}^N\sum_{k=1}^K \Vert X(n,k) - \hat X(n,k) \Vert^2
\end{equation}
where $N$ is the number of TF unit, $X$ and $\hat X$ represent the short-time Fourier transform (STFT) of clean and estimated target speech, respectively.



\section{EXPERIMENTS}
\label{sec:experiment}


\subsection{Dataset}
\label{ssec:ataset}

The proposed system is evaluated by the WSJ0-2mix datasets\footnote{Available at: http://www.merl.com/demos/deep-clustering}. In WSJ0-2mix, each sentence contains two speakers. 
The WSJ0-2mix dataset introduced in  \cite{hershey2016deep} is derived from the WSJ0 corpus \cite{garofalo2007csr}. 
The 30h training set and the 10h validation set contain two-speaker mixtures generated by randomly selecting from 49 male and 51 female speakers from si\_tr\_s. 
The Signal-to-Noise Ratios (SNRs) are uniformly chosen between 0 dB and 5 dB. 
The 5h test set is generated similarly by using utterances from 16 speakers from  si\_et\_05, which do not appear in the training and validation sets. 
The test set includes 1603 F\&M sentences, 867 M\&M sentences, and 530 F\&F sentences.

For each mixture, we randomly choose an anchor utterance from the target speaker (different from the utterance in the mixture).

\subsection{Baseline model and metrics setting}
We compare the proposed algorithm with two baselines: SpeakerBeam+DC\cite{vzmolikova2019speakerbeam} and Speakerfilter \cite{he2020speakerfilter}.

Compared with the Speakerfilter, the proposed method has two improvements: 1) the complex spectrum, not the magnitude spectrum, is used in the GCNN part, which can characterize the target speaker better. 2), The time-domain WaveUNet is used in the beginning and ending, which can separate the target speaker better. The proposed method also can be referred to as Speakerfilter+RI+WaveUNet. Based on this, we set up two comparison methods to illustrate the improvement's impact on the separation effect: Speakerfilter+RI and Speakerfilter+WaveUNet.


The performance is evaluated with two objective metrics: perceptual evaluation of speech quality (PESQ) \cite{rix2001perceptual} and Signal-to-Distortion Ratio (SDR) \cite{vincent2006performance}. 
For both of the PESQ and SDR metrics, the higher number indicates better performance.

\subsection{Evaluation results}
\label{ssec:Result}

\begin{table}[h]
	\renewcommand\arraystretch{1.0}
	\caption{Average SDR score on the test set for different methods.}
	\begin{minipage}[b]{1.0\linewidth}	
    \end{minipage}
	\centering
	\begin{tabular}{cc}
		\hline
		Method                          & SDR  \\ \hline
		unprocessed &2.62 \\ \hline
		SpeakerBeam+DC \cite{vzmolikova2019speakerbeam}  & 13.50 \\ \hline
		Speakerfilter \cite{he2020speakerfilter}  & 13.89 \\ \hline
		Proposed    & \textbf{14.95} \\ \hline
	\end{tabular}
	\label{tab:result_sdr}
\end{table}

Table \ref{tab:result_sdr} presents the comprehensive evaluation for different approaches on SDR. It can be seen that the proposed method obtains the best performance. The baseline models are using the T-F domain only for target speaker extraction. However, the proposed method leverages the time-domain modeling ability of the WaveUNet and the frequency-domain modeling ability of the GCRN, simultaneously, that can obtain a better performance. Compared with the unprocessed, the proposed method improves the SDR nearly 12.33 dB.

\begin{table}[h]
	\renewcommand\arraystretch{1.1}
	\caption{The SDR performance under different modifications on Speakerfilter.}
	\centering
	\begin{tabular}{p{3.7cm}p{0.6cm}p{0.6cm}p{0.6cm}p{1.3cm}}
		\hline
		& F\&M & F\&F & M\&M & Average \\ \hline
		\small unprocessed				  & 2.58    & 2.71   & 2.65   & 2.62      \\ \hline
		\small Speakerfilter             & 15.80   & 11.80  & 11.64  & 13.89      \\ \hline
		\small Speakerfilter+RI                       & 15.96   & 11.72  & 11.78  & 13.86      \\ \hline
		\small Speakerfilter+WaveUNet      & 16.16   & 12.84  & 12.67  & 14.26      \\ \hline
		\small Speakerfilter+RI+WaveUNet       & \textbf{16.52}   & \textbf{12.91}  & \textbf{13.68}  & \textbf{14.95}      \\ \hline
	\end{tabular}
	\label{tab:diff_conf_sdr}
\end{table}

\begin{table}[h]
	\renewcommand\arraystretch{1.0}
	\caption{The PESQ performance under different modifications on Speakerfilter.}
	\centering
	\begin{tabular}{p{3.7cm}p{0.6cm}p{0.6cm}p{0.6cm}p{1.3cm}}
		\hline
		& F\&M & F\&F & M\&M & Average \\ \hline
		\small unprocessed					   & 2.15   & 2.13   & 2.25   & 2.17      \\ \hline
		\small Speakerfilter                  & 3.30   & 2.90   & 2.91   & 3.12      \\ \hline
		\small Speakerfilter+RI                             & 3.26   & 2.87   & 2.91   & 3.09      \\ \hline
		\small Speakerfilter+WaveUNet                       & 3.32   & \textbf{3.01}   & 2.97   & 3.17     \\ \hline
		\small Speakerfilter+RI+WaveUNet           & \textbf{3.37}   & \textbf{3.01}   & \textbf{3.12}   & \textbf{3.24}      \\ \hline
	\end{tabular}
		\label{tab:diff_conf_pesq}
\end{table}

As shown in Table \ref{tab:diff_conf_sdr} and \ref{tab:diff_conf_pesq}, most of the modifications have positive effect on the baseline. On average, the Speaker+RI+WaveUNet model obtains the best results. 

The WavUNet has a large improvement for Speakerfilter and Speakerfilter+RI models. The average SDR improves 0.37 (from 13.89 to 14.26) and 1.09 (from 13.86 to 14.95) dB, respectively. The average PESQ improves 0.05 (from 3.12 to 3.17) and 0.15 (from 3.09 to 3.24) dB, respectively. We should note that, after adding the WaveUNet module, the RI-based model ($\triangle SDR=1.09, \triangle PESQ=0.15$) is more effective than the amplitude spectrum-based model ($\triangle SDR=0.37, \triangle PESQ=0.05$).

For the Speakerfilter system, after changing the magnitude spectrum to the complex spectrum, both the SDR and the PESQ are reduced by 0.03 dB. However, in the Speakerfilter+WaveUNet system, after changing the magnitude spectrum to the complex spectrum, the SDR and the PESQ are increased by 0.69 and 0.07 dB, respectively, which means that the complex spectrum feature can boost the speaker extractor performance with the participation of WaveUNet.

As mentioned above, the complex spectrum feature and the WaveUNet can boost each other.

\begin{figure}[h]
\centering
\subfigure[Speakerfilter result]{ 
\begin{minipage}{3.5cm}
\centering 
\includegraphics[width=3.5cm]{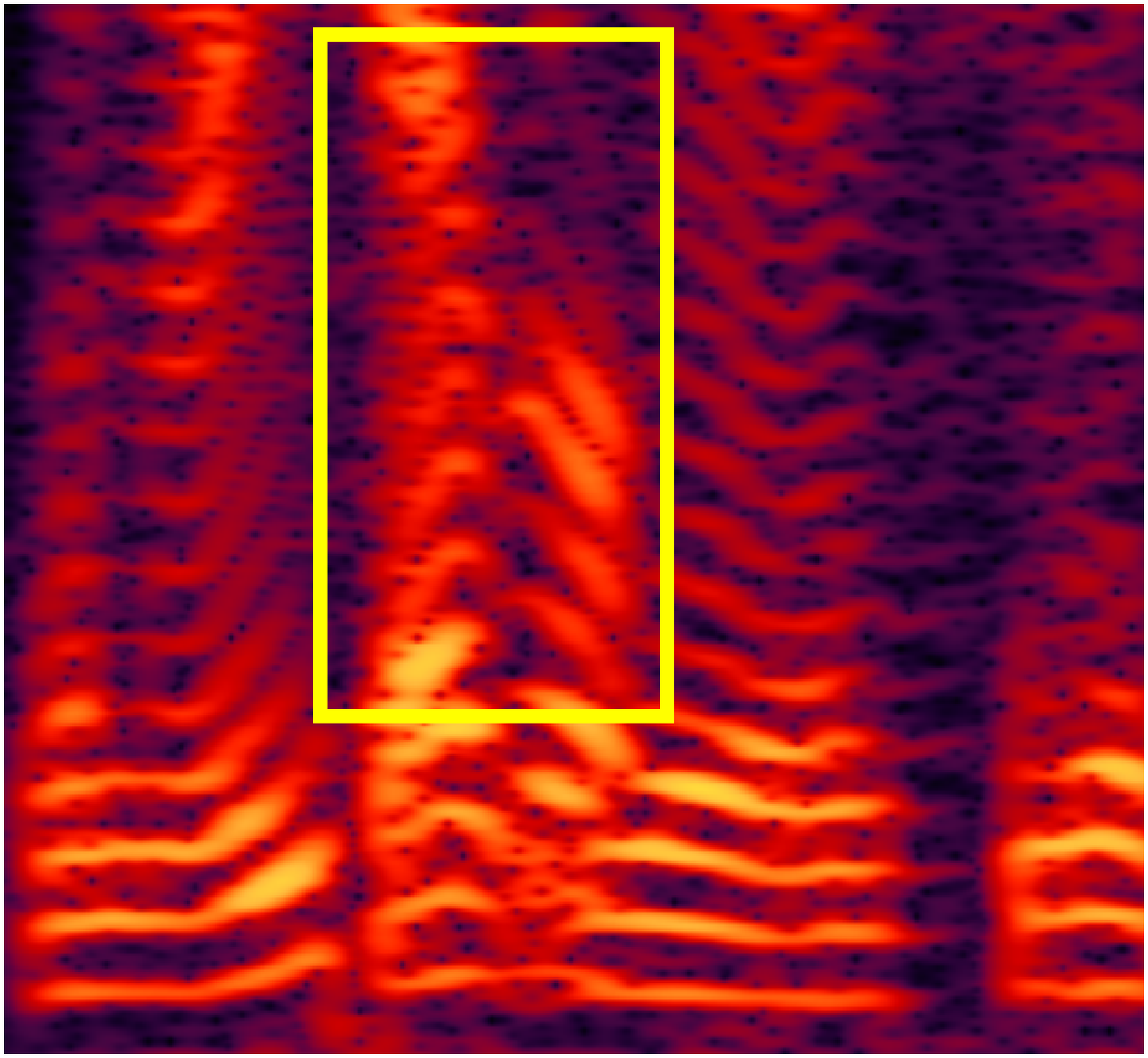}
\end{minipage}
}
\subfigure[Speakerfilter-Pro result]{ 
\begin{minipage}{3.5cm}
\centering 
\includegraphics[width=3.5cm]{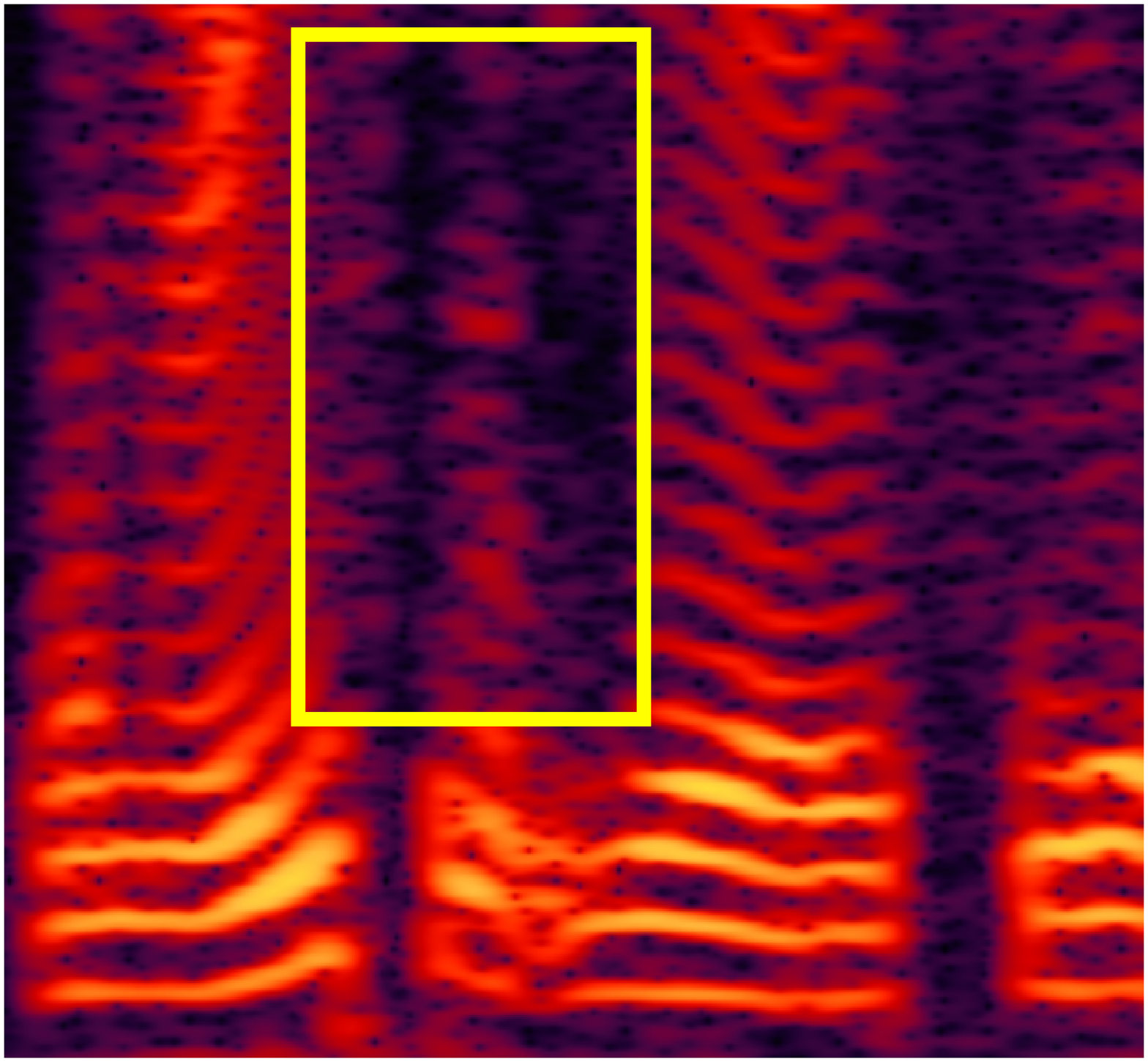}
\end{minipage}
}
\caption{Spectrograms of extracted speech using Speakerfilter and Speakerfilter-Pro. (a) The spectrogram of the separated target speech using Speakerfilter. (b) The spectrogram of the separated target speech using the Spekaerfilter-Pro.}
\label{fig:Result}
\end{figure}
Fig. 3 illustrates the spectrogram of separated speech using Speakerfilter and Speakerfilter-Pro, respectively. Our method has better performance on target speaker separation. As shown in the yellow rectangle, the proposed method can suppress the interfering speaker better than Speakerfilter.

\section{Conclusions}
\label{sec:conclusions}

This paper proposes a target speaker extractor, where WaveUNet, CRN, and BGRU-GCNN are used for time-domain feature enhancer, speech separator, and speaker feature extractor, respectively.
By leveraging the time-domain modeling ability of WaveUNet and the frequency-domain modeling ability of the CRN, the target speaker can be tracked well. According to the experiment results, the proposed method achieves better performance than other baseline methods. In the future, we will explore the robustness problem in the presence of noise and reverberation.

\section{Acknowledgments}
\label{sec:majhead}

This research was partly supported by the China National Nature Science Foundation (No. 61876214).

\newpage
\bibliographystyle{IEEEbib}
\bibliography{strings,refs}

\end{document}